\documentclass[prd, aps, nofootinbib,preprintnumbers, showpacs, superscriptaddress]{revtex4}

\usepackage{latexsym}
\usepackage{amssymb}
\usepackage{amsfonts}
\usepackage{amsmath}
\usepackage{bm}
\usepackage[dvips]{graphicx}
\usepackage{color}
\usepackage{times}

\allowdisplaybreaks

\begin{document}

\title{The horizon problem as a clue: a smooth big bang?}

\date{\today}
\label{firstpage}

\author{Carles Bona}
\affiliation{
Universitat Illes Balears, Institute for Computational Applications with Community Code (IAC3), Palma 07122, Spain
}
\affiliation{
 Institute for Space Studies of Catalonia, Barcelona 08034, Spain
}

\begin{abstract}
The necessary and sufficient condition for the absence of particle horizons in a big bang Friedmann-Robertson-Walker universe is provided. It happens to be a "smooth big bang" initial condition: the proper time derivative of the expansion factor $a(t)$ must be finite at the big bang. Equivalently, the energy density must not diverge faster than $a^{-2}$ at the big bang. This is just an initial condition: only the scale factor asymptotic behavior at the very moment of the big bang matters. The causal connection between remote space regions could then take place immediately after the big bang. Even $10^{-36}$ seconds of proper time, would allow for an infinite number of light crossing times between any two space regions, no matter how far apart. This justifies inflationary scenarios starting from a quasi-homogeneous scalar field close to equilibrium. The high degree of homogeneity of the Cosmological Background Radiation can be seen then not as a problem, but rather as a clue to the equation of state in the big bang limit.
\end{abstract}

\pacs{
98.80.Jk  
}
\maketitle


\section{Introduction.}

The Cosmological Principle postulates the space homogeneity and isotropy of the Universe. This basic assumption is confirmed by the cosmic microwave background (CMB) observations~\cite{COBE, WMAP, Planck_2013, Planck_2015}. This can be taken just as a basic mathematical axiom, which would not require any justification, or rather as a physically relevant statement, implying the existence of some mechanism leading to homogeneity.

The CMB data show a remarkable isotropy (to one part in $10^5$) for the electromagnetic radiation coming from space regions that where very far apart at the time of emission, which confirms homogeneity as well. The CMB origin can be dated back to the decoupling time, where the Universe became transparent, quite close (about just 380.000 years) to the big bang. This could be a problem if different emission points where causally disconnected at the decoupling time (cosmic horizon problem)~\cite{Rindler, Weinberg, Misner}, as it would be hard to explain the high degree of homogeneity without any causal interaction (homogeneity problem), unless one is ready to assume a highly fine-tuned initial condition.

Of course, the horizon problem would not be there if one assumes an eternal Universe, with no big bang~\cite{Starobinskky, Ellis_Maartens}. There would be plenty of time for the Universe to reach an homogeneous (at cosmic scales) matter distribution (see refs.~\cite{3+2_paper, 4+1_paper} for a simple way of building singularity-free models).

But the mainstream solution for the horizon problem is the inflation paradigm~\cite{Guth, Linde}, which encloses many different models (see ref.~\cite{Cosmology_review} for a recent review). In some of them, the early Universe is assumed to be vacuum dominated (quantum vacuum energy can be described by a cosmological constant in classical terms). In some others, the basic assumption is the existence of some cosmic scalar field $\Phi$ (the "inflaton") in a local equilibrium/quasiequilibrium state. Ordinary matter and radiation arise a bit later, as a consequence of the energy liberated by a phase transition. In this way, the Universe experiments a huge acceleration, that manages to place the cosmic horizons beyond the limits of the region we can observe today.

This is just a partial implementation of the Cosmological Principle, as the resulting Universe is just piecewise homogeneous (although in very big pieces). Moreover, inflation is supposed to start at some non-zero time. The cosmological constant solution for the expansion factor $a(t)$ zero space curvature case is actually
\begin{equation}\label{expansion 0}
    a(t) \sim e^{\sqrt{\Lambda/3}\,ct} \,,
\end{equation}
which cannot reach the big bang singularity ($a=0$). There are some physical arguments for such delay~\cite{Guth, Linde}, because the energy at very early times (about $10^{-36}$ seconds) was much beyond our current accelerators capabilities, even beyond the validity of the classical gravitation approach.

A General Relativity complete solution, starting from the big bang singularity, must then give an answer to the questions arising from most inflationary models assumptions~\cite{Guth, Linde}. In order to be specific, let me choose the
scalar field approach (the details could be adapted to other cases), in which the energy density is given by
\begin{equation}\label{energy scalar}
    \rho = \frac{1}{2}\dot{\Phi}^2 + \frac{1}{2}(\nabla\,\Phi)^2 + V(\Phi)\,.
\end{equation}
The basic assumption is that inflation starts from an initial equilibrium or quasi-equilibrium state of the inflaton field, that is
\begin{equation}\label{ansatz}
    \dot{\Phi}^2 +  (\nabla\,\Phi)^2 \ll V(\Phi)\,,
\end{equation}
so that the inflaton potential $V$ mimics a cosmological constant, leading to an exponential expansion rate~\footnote{Note that the origin of any potential is arbitrary, so that the value of potential term in equation (\ref{ansatz}) should be understood as the magnitude of the relevant potential barrier}.

How can this happen in just $10^{-36}$ seconds? How can this degree of space and time homogeneity be reached everywhere so fast without fine-tuned initial conditions?.

This paper will show a precise answer for this problem: the absence of particle horizons. The necessary and sufficient condition for this to happen in a big bang Friedmann-Robertson-Walker (FRW) universe is just a "smooth big bang" initial condition: the proper time derivative of the expansion factor $a(t)$ must be finite at the initial singularity ($a=0$). This condition is yet satisfied by the cosmological constant solution of negative curvature, namely:
\begin{equation}\label{expansion plus minus}
    a(t) \sim \sinh(\sqrt{\Lambda/3}\,ct)\,,
\end{equation}
which could then be prolonged up to the beginning of time without any horizon problem.

Of course, the physics details of that early pre-inflation era are beyond reach, because physics at such a high energy range is much beyond current experimental capabilities. This is why I will rather follow here Synge's g-method (runninig fiel equations backwards)~\cite{Synge}: take the required geometrical properties (absence of horizons) as the starting point and then deduce some constraint on the unknown matter/energy behaviour.

\section{FRW causal structure.}

In General Relativity, the Cosmological Principle can be translated into precise mathematical terms: the Universe would be described by the FRW family of line elements:
\begin{equation}\label{FRW}
    ds^2 = - c^2\, dt^2 + a^2(t)\,\gamma_{ij}\, dx^i\,dx^j,
\end{equation}
where the time-independent metric $\gamma_{ij}$ can be written in spherical coordinates as
\begin{equation}
    \gamma_{ij}\, dx^i dx^j = \frac{dr^2}{1-k r^2} + r^2(d\theta^2 + sin^2\theta\, d\varphi^2)\;\;(k=0,\pm 1).
\end{equation}

Let me start the analysis by switching from the proper time coordinate $t$ to the conformal time coordinate defined by
\begin{equation}\label{timetransform}
    \eta\,\equiv\, \int \frac{dt}{a(t)}\,,
\end{equation}
so one can visualize that FRW metrics are just proportional to a static metric, namely
\begin{equation}\label{conformal}
    ds^2 = a^2(\eta)\,(- c^2\, d\eta^2 +\,\gamma_{ij}\, dx^i\,dx^j) \equiv a^2(\eta)\,d\sigma^2 \,.
\end{equation}
This is a crucial point, because conformally related metrics share the same light-cone structure, that is
\begin{equation}
    ds^2 = 0 \;\Leftrightarrow\; d\sigma^2=0\,.
\end{equation}

\begin{figure}[h]
\centering
\includegraphics[width=6cm,height=6cm]{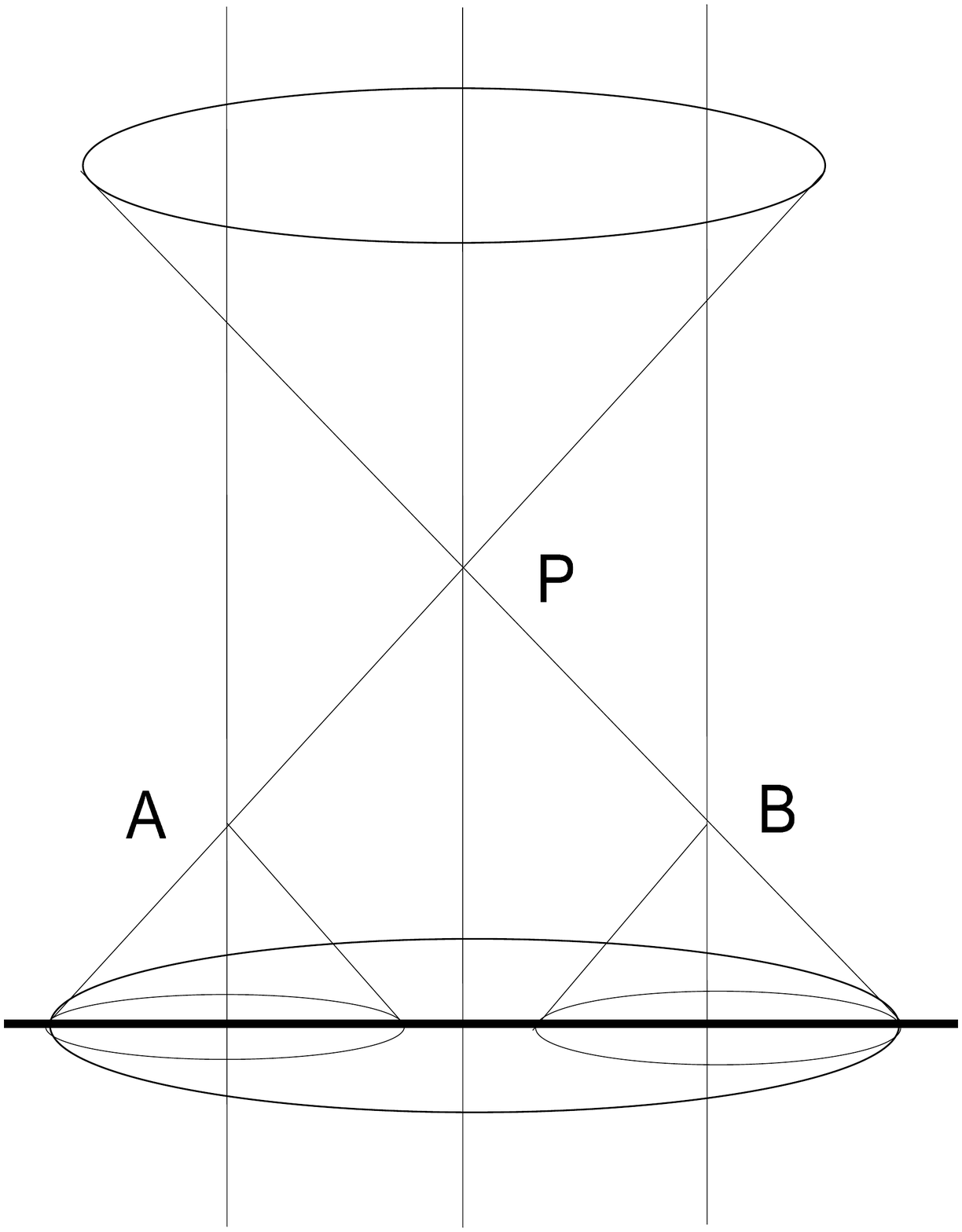}
\caption{Light cones of the spatially flat FRW spacetime (same as for Minkowski). Vertical lines describe the cosmological (comoving) observers trajectories. These worldlines are always inside the local light cones. The thick horizontal line marks the big bang instant, assuming that it corresponds to a finite value of the conformal time.}
\label{cones}
\end{figure}

This simple argument shows that the light cone structure of FRW universes is static. In order to be more specific, let me restrict for a while to the spatially flat ($k=0$) case, where the conformal static metric is just flat Minkowski space. The light cone structure is displayed in figure~\ref{cones}. Vertical lines correspond to the worldlines of comoving cosmological observers, both in the original FRW and in the conformal Minkowski spacetime. These world lines are always inside their local light cones.

Figure~\ref{cones} shows also the big bang instant (thick line), assuming that it happened at a finite value of the conformal time. The part below the big bang line must then be removed from the diagram, so you do not have the whole Minkowski causal structure: only the upper half of it. This means that one can be receiving light signals, from opposite space directions, emitted by events ($A$ and $B$ in the figure) that are causally disconnected: their past light cones do not intersect because they are cut by the big bang. These are termed 'particle horizons' or 'cosmic horizons'.

\begin{figure}[h]
\centering
\includegraphics[width=6cm,height=6cm]{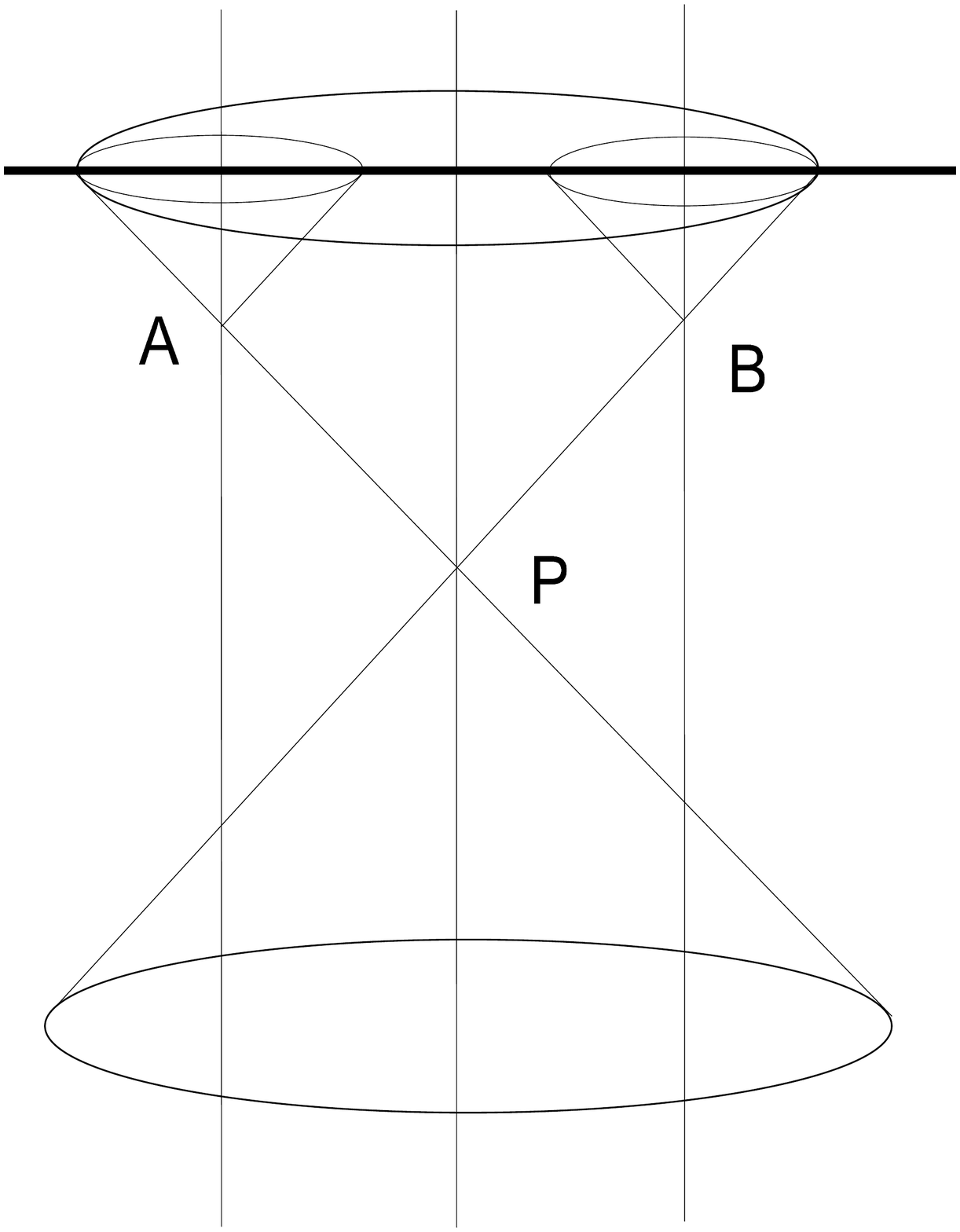}
\caption{Same as figure~\ref{cones}, but assuming now a smooth big bang at $t=0$, which corresponds then to $\eta\rightarrow -\infty$ in conformal time (thick horizontal line). No particle horizons arise, as an infinite number of crossing times can take place, no matter how far a part the interacting regions are. Late time accelerated expansion is also taken into account (see text): proper time infinity corresponds then to a finite value of conformal time, leading to the formation of event horizons.}
\label{inverted_cones}
\end{figure}

Note, however, that this is just a naive first approximation, as no acceleration effects are accounted for yet. The late time accelerated expansion of the Universe (\ref{expansion 0}) leads to a finite value of the improper integral
\begin{equation}\label{time_lapse}
    \triangle \eta = \int_{t_0}^{\infty} \frac{dt'}{a(t')}\,,
\end{equation}
meaning that (proper time) future infinity will actually correspond to a finite value of conformal time, as shown in  Figure~\ref{inverted_cones} (upper part), and event horizons appear as a result.

Let us come back, however, to the cosmic horizon issue (early times era). The relevant integral
\begin{equation}\label{timetlapse_initial}
    \triangle \eta = \int_{0}^{t_0} \frac{dt'}{a(t')}\,,
\end{equation}
is also improper. The result can then be either finite or infinite, depending on the scale factor asymptotic behaviour at the big bang. If it is finite, the big bang will occur at a finite value of conformal time, as depicted in figure~1, and particle horizons will arise as a result. On the contrary, if the asymptotic behavior is given by
\begin{equation}\label{asymptotics}
a(t)\,\simeq\, t^n\;\;\;\;(n\geq 1)\,,
\end{equation}
the improper integral (\ref{timetlapse_initial}) will diverge, meaning that the big bang would arise for $\eta \rightarrow -\infty$ conformal time. The thick line in Figure~1 is then removed in Figure~\ref{inverted_cones}, where the lower half corresponds actually to Minkowski space.

It follows that any pair of events (like $A$ and $B$ in the second figure) are now causally connected: an infinite number of light crossing times could have occurred since the big bang. There are no particle horizons, and the high degree of homogeneity in distant regions can be explained by eventual interactions in the past (infinite crossing times available). This result is independent of the sign of the space curvature.

Note that the proper time chronology is not affected in any way. Only the scale factor asymptotic behavior at the very moment of the big bang matters. No matter how small fraction of proper time has elapsed since the big bang (even $10^{-36}$ seconds, before the assumed beginning of inflation) would still correspond to an infinite amount of conformal time, allowing infinite crossing times as a result. The quasi-homogeneity condition (\ref{ansatz}) is a quite natural assumption in this scenario.

Condition (\ref{asymptotics}) has a direct implication on the matter/energy content. Taking a time derivative, one gets
\begin{equation}\label{initial}
\dot{a}(t)|_{a=0}\,<\, \infty\,,
\end{equation}
meaning that the proper time slope of the expansion factor is finite at the big bang. This "smooth big bang" as the beginning of our Universe is an extremely simple hypothesis that ensures the absence of any particle horizon whatsoever. Allowing for the Friedmann equation
\begin{equation}\label{Friedmann}
   \dot{a}^{2} = \frac{\rho}{3}a^2-k\,,
\end{equation}
the smooth big bang condition can be expressed alternatively in terms of the energy density $\rho$, that is
\begin{equation}\label{density}
(\rho\,a^2)|_{a=0}\,<\, \infty\,.
\end{equation}
This of course excludes the presence of ordinary matter and/or radiation at the very big bang moment (in keeping with the inflation chronology). On the contrary, the negative curvature brand of the cosmological constant solution (\ref{expansion plus minus}) provides the simplest viable option.


\section{Conclusions and outlook.}

The high degree of homogeneity of the CMB can be seen then not as a problem, but rather as a clue to the equation of state at the very moment of the big bang, which probably could not be deduced by other means, as we cannot be sure that our current physics understanding can be extrapolated that far.

In this sense, the smooth big bang condition (\ref{initial}) has been identified as the necessary and sufficient conditions for the absence of particle horizons. One can interpret this result in two different ways:
\begin{enumerate}
  \item As a natural way for justifying the inflaton smoothness condition (\ref{ansatz}), then supporting the inflation approach. This support is reinforced by the fact that, allowing for the equivalent initial condition (\ref{density}), ordinary matter and radiation could not be present at the very big bang moment, in keeping with inflation chronology.
  \item As opening an avenue to alternative approaches, as no particle horizons are formed. Condition (\ref{density}) can actually be taken as the starting point. One can guess some initial density evolution profile $\rho(a)$ and then, using the Friedmann equation (\ref{Friedmann}), express the resulting FRW line element in fully explicit form:
      \begin{equation*}
      \;\;\;   ds^2 = -\left( \frac{\rho}{3}\,a^2 - k\right)^{-1} da^2 + a^2\,\gamma_{ij}\, dx^i\,dx^j\,,
      \end{equation*}
      where the expansion factor is used here as a time coordinate~\cite{4+1_paper}. If (and only if) the energy density verifies (\ref{density}), the resulting metric will be horizon-free, so that the CMB homogeneity will not be a problem at all. Of course, the flatness problem remains, so any alternative to inflation should address this issue.
\end{enumerate}

As a final remark, let me briefly comment on the plausibility of the smooth big bang initial condition (\ref{initial}). Measuring the probability of specific initial conditions is actually an open issue~\cite{Cosmology_review}. If one assumes that (\ref{initial}) does not hold, meaning that the initial slope is infinite, then this infinite slope can only decrease, so one would get an initial decelerating scenario. This point can be illustrated with the simple power-law asymptotic behavior (\ref{asymptotics}):
\begin{equation}\label{acceleration}
n > 1 \;\Rightarrow\; \ddot{a} < 0\,.
\end{equation}
This means that assuming an initial acceleration, which is quite a natural assumption in inflationary scenarios, is a sufficient condition (although not a necessary one) for a smooth big bang.

\section*{Acknowledgement.}

The author acknowledges support from the Spanish Ministry of Economy, Industry and Competitiveness grants AYA2016-80289-P and AYA2017-82089-ERC (AEI/FEDER, European Union).




%
%

\bibliographystyle{prsty}

\end{document}